\documentclass[aps,prl,twocolumn,floatfix,superscriptaddress]{revtex4-2}

\usepackage{amssymb,amsmath,amstext}                
\usepackage{graphicx}                                               
\usepackage{epstopdf}                                               
\usepackage{color} 
\usepackage[dvipsnames]{xcolor}
\usepackage{bm}                                                        
\usepackage{verbatim}
\usepackage{appendix}                                              
\usepackage[utf8]{inputenc}
\usepackage{bbold}
\usepackage{bbm}
\usepackage{ulem}
\usepackage{braket}
\usepackage{latexsym}
\usepackage[colorlinks=true,citecolor=blue,linkcolor=magenta]{hyperref}

\newcommand{\vect}[1]{\mathbf{#1}}

\def\be{\begin{equation}}
\def\ee{\end{equation}}
\def\bea{\begin{eqnarray}}
\def\eea{\end{eqnarray}}

\def\bi{\begin{itemize}}
\def\ei{\end{itemize}}
\def\ben{\begin{enumerate}}
\def\een{\end{enumerate}}

\newcommand{\changed}[1]{{\color{black}{#1}}}

\begin{document}

\title {Towards Timetronics with Photonic Systems}

\author{Ali Emami Kopaei}
\thanks{ali.emami.app@gmail.com}
\affiliation{Szkoła Doktorska Nauk Ścisłych i Przyrodniczych, Wydział Fizyki, Astronomii i Informatyki Stosowanej, Uniwersytet Jagiello\'nski, ulica Profesora Stanisława Łojasiewicza 11, PL-30-348 Kraków, Poland}
\affiliation{Instytut Fizyki Teoretycznej, Wydział Fizyki, Astronomii i Informatyki Stosowanej, Uniwersytet Jagiello\'nski, ulica Profesora Stanisława Łojasiewicza 11, PL-30-348 Kraków, Poland}
\author{Karthik Subramaniam Eswaran}
\affiliation{Szkoła Doktorska Nauk Ścisłych i Przyrodniczych, Wydział Fizyki, Astronomii i Informatyki Stosowanej, Uniwersytet Jagiello\'nski, ulica Profesora Stanisława Łojasiewicza 11, PL-30-348 Kraków, Poland}
\affiliation{Instytut Fizyki Teoretycznej, Wydział Fizyki, Astronomii i Informatyki Stosowanej, Uniwersytet Jagiello\'nski, ulica Profesora Stanisława Łojasiewicza 11, PL-30-348 Kraków, Poland}
\author{Arkadiusz Kosior}
\affiliation{Institut f\"ur Theoretische Physik, Universit\"at Innsbruck, A-6020 Innsbruck, Austria}
\author{Daniel Hodgson}
\affiliation{School of Physics and Astronomy, University of Leeds, Leeds, UK, LS2 9JT}
\author{Andrey Matsko}
\affiliation{Jet Propulsion Laboratory, California Institute of Technology, 4800 Oak Grove Drive, Pasadena, California 91109-8099, USA}
\author{Hossein Taheri}
\affiliation{Department of Electrical and Computer Engineering, University of California Riverside, 3401 Watkins Drive, Riverside, CA 92521}
\author{ Almut Beige}
\affiliation{School of Physics and Astronomy, University of Leeds, Leeds, UK, LS2 9JT}
\author{Krzysztof Sacha}
\thanks{krzysztof.sacha@uj.edu.pl}
\affiliation{Instytut Fizyki Teoretycznej, Wydział Fizyki, Astronomii i Informatyki Stosowanej, Uniwersytet Jagiello\'nski, ulica Profesora Stanisława Łojasiewicza 11, PL-30-348 Kraków, Poland}
\affiliation{Centrum Marka Kaca, Uniwersytet Jagiello\'nski, ulica Profesora Stanisława Łojasiewicza 11, PL-30-348 Kraków, Poland}

\begin{abstract}
Periodic driving of \changed{particles} can create crystalline structures in \changed{their dynamics}. Such systems can be used to study solid-state physics phenomena in the time domain. In addition, it is possible to \changed{realize photonic time crystals and to engineer the wave-number band structure  of optical devices by periodic temporal modulation of the properties of light-propagating media}. \changed{Here we introduce a versatile approach which uses traveling wave resonators to emulate} various condensed matter phases in the time dimension.
This is achieved by utilizing temporal modulation of \changed{the permittivity and the shape of small segments of the resonators}. The required frequency and depth of the modulation are experimentally achievable \changed{which opens a pathway for} the practical realisation of crystalline structures in time \changed{in} microwave and \changed{in} optical systems.
\end{abstract}

\date{\today}

\maketitle

In periodically driven atomic and solid-state systems, as well as in nonlinear optical systems, it is possible to realize discrete time crystals that spontaneously break discrete translational symmetry in time and begin to evolve with a period longer than that dictated by the periodic perturbation \cite{Sacha2015,Khemani16,ElseFTC,Zhang2017, Choi2017,Pal2018,Rovny2018, Smits2018,Mi2022,Randall2021,Frey2022, Kessler2020,Kyprianidis2021, Xu2021,Taheri2020,Taheri2022a, Taheri2022b, Liu2023,Bao2024,Kazuya2024,Liu2024,Liu2024a,kopaei2023}.  New periodic evolution forms spontaneously, creating new crystalline structures in time. Periodically perturbed atomic systems are also well-suited for realizing a wide range of phases known from condensed matter physics, but observed in the time dimension \cite{SachaTC2020,GuoBook2021,Hannaford2022}, such as Anderson and many-body localization \cite{sacha16,Mierzejewski2017}, Mott and topological insulators \cite{Sacha15a,Giergiel2018b}, fractonic excitations \cite{Giergiel2022}, as well as higher-dimensional topological systems \cite{Giergiel2021,Braver2023}.  The flexibility of controlling and modifying various solid-state physics behaviors in time through periodic perturbation control suggests practical applications. Analogous to electronics, timetronics concerns the research and design of potentially useful devices where crystalline structures in time play a key role \cite{Giergiel2024}.

\changed{Now, let us consider} electromagnetic waves \changed{propagating} in media \changed{whose} refractive index changes periodically in space or \changed{in} time \cite{Galiffi2022,Pendry24,67cae99304e54fd0aa49df46cdcba5c2}. 
In the first case, \changed{one can observe} photonic crystals in space which exhibit a band structure in the frequency domain.  In the second case, photonic time crystals \changed{emerge, and} the wave number domain reveals a band structure. In the optical regime, the experimental realization of photonic time crystals is a formidable challenge because the required modulation depth of the refractive index must be significant. \changed{In addition,} the frequency of \changed{refractive index changes must be} comparable to optical frequencies \cite{Lustig:23}.

In this Letter, we pave the way for {\it optical timetronics} by demonstrating that simple traveling wave \changed{resonators} can exhibit a broad range of condensed matter phases, \changed{including}  combinations of different phases, in the time domain. For example, Anderson or topological insulators can be realized \changed{and} different behaviors can be connected together. \changed{Moreover, experiments can be controlled by external fields which can be present during a certain phase or during the entire experiment, and which can be completely reconfigured at any moment.} All of this is possible through time periodic modulation of the permittivity adapted to the shape of \changed{small fragments} of the \changed{resonators} where the modulation is performed. That is, temporal harmonics of the modulation match resonantly with spatial harmonics of the \changed{resonator fragments} and determine the effective behavior of the system. The \changed{proposed} time modulation  is experimentally feasible and allows for processing electromagnetic signals using phenomena known in condensed matter physics, thus \changed{introducing a path towards} timetronics, i.e., applications where crystalline structures in time play a crucial role \cite{Giergiel2024}. 

\changed{In the following,} we present our idea by providing a detailed prescription for designing an arbitrary one-dimensional band structure. We then present two specific examples: the Su–Schrieffer–Heeger (SSH) model \cite{Su1979,Asboth2016short} and the Wannier-Stark ladder \cite{Wannier_1962,Philipp_2015}. \changed{In \cite{SM}, we describe how to generalize our idea to two-dimensional condensed matter phases in periodically modulated coupled traveling-wave resonators.} 
 
\begin{figure}[t]
\centering
\includegraphics[width=0.99\columnwidth]{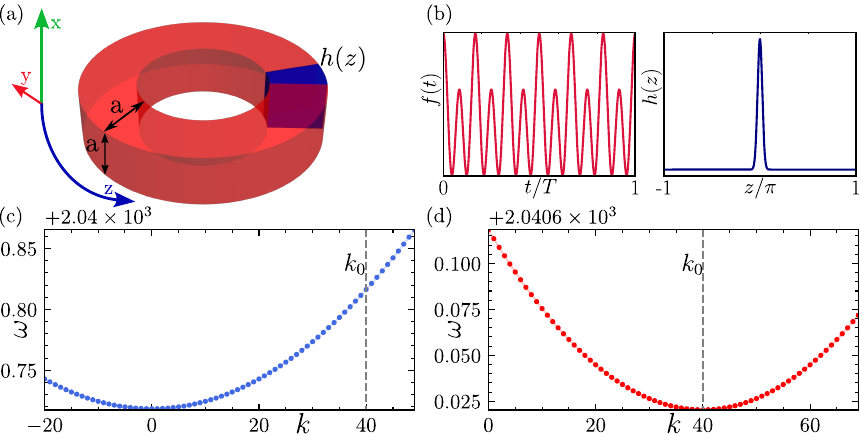}
\caption{
(a) Resonator in the form of a closed ring with a square cross-section. The circumference of the ring, in the units used in the text, is $2\pi$. In a small segment of the resonator described by $h(z)$, the permittivity is periodically modulated in time with the frequency $\Omega=2\pi/T$. (b) Example of the periodic modulation, $f(t)$, of the permittivity (left) in a small fragment of the resonator described by $h(z)$ of the Gaussian shape (right). This example is analyzed in the text. (c) Dispersion relation (in the laboratory frame) for longitudinal modes of the resonator with the length of the side of the cross-section area of the resonator \changed{ $a=0.014\pi$.} Points indicate the discrete spectrum of the ring-shaped resonator. (d) In the frame moving with the frequency $\Omega$ which matches the free spectral range of the resonator, the dispersion relation for the longitudinal modes has a minimum at $k_{0}=40$, and superpositions of waves with $k \approx k_{0}$ evolve extremely slowly.}
\label{fig:1}
\end{figure}

 As an illustration of our idea, let us consider a closed resonator in the form of a ring with a square cross-section [Fig.~\ref{fig:1}(a)].
For simplicity, we re-scale the magnetic field vector, $\vect{H} \rightarrow \sqrt{\varepsilon_0 / \mu_0} \, \vect{H}$, and use $L / 2\pi$ and $L / 2\pi c$ as the units of length and time, respectively, where $L$ is the circumference of the ring, $c$ is the speed of light, and $\mu_0$ and $\varepsilon_0$ are the vacuum permeability and permittivity \cite{SM}. Most importantly, we assume that the relative permittivity of a material $\varepsilon$ is constant everywhere in the resonator except for a small segment where $\varepsilon$ changes periodically in time, \changed{for example such that}
\be
\varepsilon(z,t)=\varepsilon_r +h(z)f(t) ~~ {\rm with} ~~ f(t+T)=f(t),
\ee
as in Fig.~\ref{fig:1}(b). Here $z$ denotes the position along the resonator and $h(z)$ and $f(t)$ are functions which vary in space and \changed{in} time respectively and describe the small segment of the resonator. In numerical calculations we set $\varepsilon_r =4$.

Next, we assume that the  TE$_{11}$ mode of the electromagnetic field has been injected into the resonator \cite{SM}. In this case, the electric field does not have a longitudinal component and we only need to consider the dependence of the transverse electric field amplitude $E(z,t)$ on time and \changed{on} space. Since $\varepsilon(z,t)$ is \changed{periodic in} time, we can apply the Floquet theorem \cite{Shirley1965,SachaTC2020} and seek solutions of Maxwell equations in the form $E(z,t)={\widetilde E}(z,t)e^{i\omega t}$ with ${\widetilde E}(z,t+T)={\widetilde E}(z,t)$. Here $\omega$ denotes the quasi-frequency of the electromagnetic field. The general solution of Maxwell equations can be obtained as a superposition of solutions ${\widetilde E}(z,t)e^{i\omega t}$. When solving Maxwell equations, it is convenient not to fully eliminate the magnetic field and to reduce them instead to \changed{the} generalized eigenvalue problem 
\be
\left[ 
\begin{array}{cc}
-i\partial_t\varepsilon-i\varepsilon\partial_t & \frac{i}{2k_\perp}\partial^2_z-ik_\perp  \\
2ik_\perp & -i\partial_t  
\end{array} 
\right]
\left[ 
\begin{array}{c}
\widetilde E \\
\widetilde H 
\end{array} 
\right]
=
\omega
\left[ 
\begin{array}{cc}
\varepsilon & 0  \\
0 & 1  
\end{array} 
\right]
\left[ 
\begin{array}{c}
\widetilde E  \\
\widetilde H 
\end{array} 
\right]
\label{exactMaxwell}
\ee
with eigenvalue $\omega$ \cite{SM}. In this equation, $H(z,t)={\widetilde H}(z,t)e^{i\omega t}$, \changed{where} ${\widetilde H}(z,t+T)={\widetilde H}(z,t)$ is the amplitude of the longitudinal component of the magnetic field and $k_\perp=\pi/a$ with $a$ being the 
length of each side of the cross-sectional area of the \changed{travelling wave} resonator [Fig.~\ref{fig:1}(a)].

\changed{When the permittivity of the resonator is constant in time (i.e., when $f(t)=0$)}, the solutions of Maxwell equations are characterized by a nonlinear dispersion relation $\omega=\sqrt{2k_\perp^2+k^2}$ where the wave number $k$ of the longitudinal modes takes integer values because we are considering 
\changed{a closed ring resonator} with periodic boundary conditions (Fig.~\ref{fig:1}). \changed{Next} let us assume that the permittivity is periodically modulated in time with \changed{ a non-zero} frequency $\Omega = 2\pi/T$ and that the resonance condition for the frequency of a wave packet \changed{circulating} around the resonator is satisfied, i.e., $\Omega$ matches the free spectral range of the resonator. \changed{In this case}, there exists a wave number $k_{0}$ for which the group velocity (calculated in the absence of the modulation) $\partial\omega/\partial k|_{k=k_0} = \Omega$. \changed{Here} the new units have been applied. For \changed{a} sufficiently long resonator, the frequency $\Omega$ of the permittivity modulation is within the experimentally achievable range, e.g., for $L=1$~cm, \changed{we have} \changed{$\Omega=150$~MHz.}

To simplify the description of the resonant behavior of the system, we will switch to a reference frame evolving with the modulation frequency $z'=z-\Omega t$ \changed{by} using the transformation $U=e^{\Omega t \partial_z}$ \cite{SM}. In the moving frame, the dispersion relation has a quadratic behavior with the minimum at the resonant wave number $k_0$, i.e. $\omega\approx \omega(k_0)+\partial^2\omega/\partial k^2|_{k=k_0}(k-k_0)^2/2$, see Fig.~\ref{fig:1}(d) \changed{and \cite{SM}}. In the present work we assume that the group velocity dispersion results from the geometry of the resonator but in general there can also be a contribution from the material properties of the resonator. In the moving frame, \changed{a superposition of waves with wave numbers $k \approx k_{0}$ evolves} very slowly, and we can average Maxwell equations over time. This is essentially the rotating wave approximation, leading to time-independent effective Maxwell equations in the moving frame \cite{Lichtenberg1992,SachaTC2020,SM}:
\be
\left[ 
\begin{array}{cc}
i\Omega\partial_z\bar\varepsilon+i\Omega\bar\varepsilon\partial_z & \frac{i}{2k_\perp}\partial^2_z-ik_\perp  \\
2ik_\perp & i\Omega\partial_z  
\end{array} 
\right]
\left[ 
\begin{array}{c}
\widetilde E \\
\widetilde H 
\end{array} 
\right]
=
\omega
\left[ 
\begin{array}{cc}
\bar\varepsilon & 0  \\
0 & 1  
\end{array} 
\right]
\left[ 
\begin{array}{c}
\widetilde E  \\
\widetilde H 
\end{array} 
\right].
\label{effMaxwell}
\ee
For clarity, we omitted the primes; \changed{all} variables and quantities refer to the moving frame. Moreover, the time-averaged permittivity $\bar\varepsilon(z)$ in the above equation can be written as
\be
\bar\varepsilon(z)=\varepsilon_r+\frac{1}{T}\int_0^T \mathrm{d}t\;h(z+\Omega t)f(t)=\varepsilon_r+\sum_mh_mf_{-m}e^{imz},
\label{avrageepsilon}
\ee
after introducing the Fourier expansions $h(z)=\sum_mh_me^{imz}$ and $f(t)=\sum_lf_le^{il\Omega t}$.

In the moving frame, the effective Maxwell equations (\ref{effMaxwell}) describe electromagnetic waves in a resonator with a time-averaged permittivity $\bar\varepsilon(z)$ which varies in space along the resonator. In the laboratory frame, the segment of the resonator in which the permittivity is modulated in time is described by a localized function $h(z)$ and, therefore, has many non-zero Fourier coefficients $h_m$. The Fourier expansion coefficients of $f(t)$ can be chosen to realize any average permittivity $\bar\varepsilon(z)$ in the effective description in the moving frame since $\bar\varepsilon(z)$ can always be expanded such that $\bar\varepsilon(z)=\varepsilon_r+\sum_m\varepsilon_me^{imz}$. All we need to do to obtain the given $\bar\varepsilon(z)$ in (\ref{avrageepsilon}) is to choose $f(t)$ such that its Fourier coefficients satisfy $f_{-m}=\varepsilon_m/h_m$. For example, we can realize $\bar\varepsilon(z)\propto \cos(sz)$ with integer $s\gg1$ and observe a band structure in the quasi-frequency, $\omega$, domain. We can realize a topological insulator or introduce disorder in a crystalline structure and realize an Anderson insulator. 

We can also combine crystalline structures with different properties, e.g., in different regions of $z$, the average permittivity $\bar\varepsilon(z)$ can reveal topologically different structures. \changed{Static electric fields, such as potential barriers or wells, can be applied or a more complex modulation of $\bar\varepsilon(z)$ can be realized.} Note that a stationary solution of (\ref{effMaxwell}) will appear as a propagating solution when we return to the laboratory frame. Thus, any condensed matter like behavior which we realize and observe versus $z$ in the moving frame will be observed in the time domain, \changed{ if we place a detector in the laboratory frame at a certain position $z_0$ along} the resonator and investigate its clicking in time. This \changed{ approach} paves the way for optical timetronics, i.e., similar to electronics, we can design time-varying systems where electromagnetic signals are processed \changed{by} employing phenomena known in solid-state physics \cite{Giergiel2024}.

\begin{figure}[t]
\centering
\includegraphics[width=0.99\columnwidth]{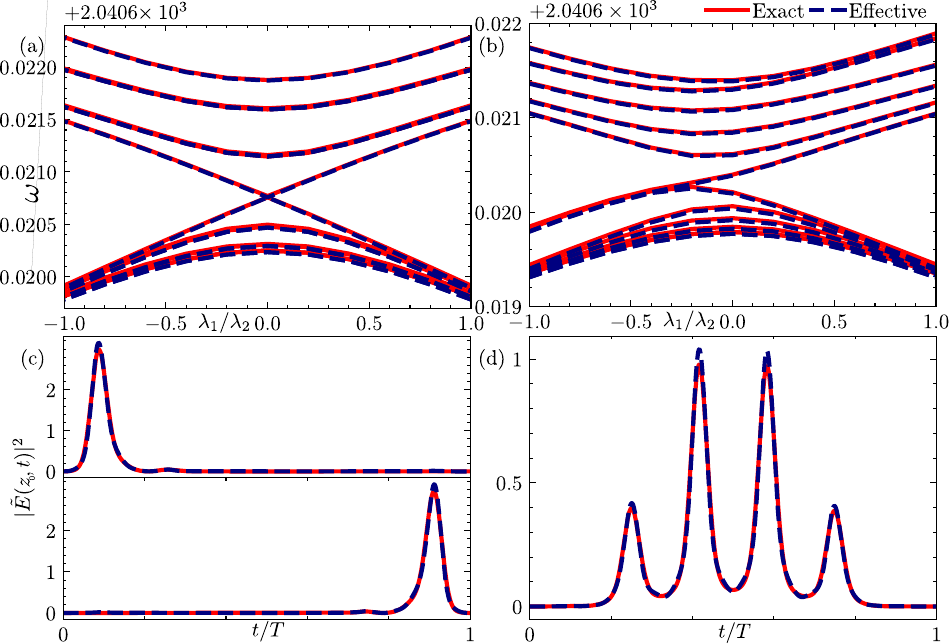}
\caption{(a) Quasi-frequency spectra corresponding to the exact and effective Maxwell equations. In the effective equations, the averaged permittivity is described by (\ref{epsilonSSH}), where $s=12$ and \changed{$\lambda_2=10^{-6}$}. (b) The same as in (a), but in the presence of additional time modulation of the permittivity, which leads to a Gaussian barrier in (\ref{epsilonSSH}) with a width of \changed{$\pi/16$} located at $z=0$. For $\lambda_1 < 0$, the formation of degenerate levels in the gap of the quasi-frequency bands can be observed. (c) Variations of the electric field at position \changed{$z_0=0$} in the laboratory frame corresponding to two quasi-frequency levels located in the gap between two bands in (b) for \changed{$\lambda_1/\lambda_2=-1$}. The field is localized around the moment in time when a Gaussian barrier in $\bar{\varepsilon}(z_0 - \Omega t)$ appears at the position $z_0$ \cite{footnote1}. (d) Similar to (c) but for a state from one of the bands. Such bulk states are delocalized along the entire period $T$. In all panels, both the exact results and the results of the effective Maxwell equations are presented.
}
\label{fig:2}
\end{figure}

As a first example, we consider a system that can reveal topologically protected edge states in the time domain \cite{Lustig2018,Giergiel2018,SachaTC2020,Hannaford2022,kopaei2022,kopaei2024}. Suppose \changed{$k_\perp=72.15$}, $h(z)=e^{-z^2/2\sigma^2}$
with $\sigma=\pi/41$ and $f(t)=(\lambda_1/h_{s/2})\cos(s\Omega t/2)+(\lambda_2/h_s)\cos(s\Omega t)$ with $s$ being even, with \changed{$\Omega=4.9\times10^{-3}$} and with $h_{s/2}$ and $h_s$ denoting the Fourier coefficients of $h(z)$, see Fig.~\ref{fig:1}(b). In the moving frame, the resulting average permittivity takes the form of a crystalline structure in space with a two-point basis,
\be
\bar\varepsilon(z)= \varepsilon_r+\lambda_1\cos(sz/2)+\lambda_2\cos(sz).
\label{epsilonSSH}
\ee
The quasi-frequency spectra obtained for $s=12$ with the help of the effective approach (\ref{effMaxwell}) and by \changed{solving} Maxwell equations (\ref{exactMaxwell}) exactly are presented in Fig.~\ref{fig:2}(a). \changed{In an experiment, the number of lattice sites can be increased either by increasing the modulation frequency $s\Omega$ or by extending the circumference of the ring resonator.} Both \changed{the effective and exact} solutions match each other very well. For example, for $\lambda_1=0$, we observe a single band in the quasi-frequency domain consisting of $s=12$ levels. When $\lambda_1$ starts to differ from zero, the crystalline structure in $\bar\varepsilon(z)$ has a two-point basis and the initially single band splits into two. When we focus on these two bands, 
the band structure of the system is equivalent to 
the well-known SSH model which reveals topologically protected edge states in the presence of edges and in the topologically nontrivial regime \cite{Su1979,Asboth2016short}. 

The crystalline structure in (\ref{epsilonSSH}) is a periodic structure in a resonator without any edge. However, by a proper additional time modulation of the permittivity we can introduce a localized barrier in the crystalline structure and thus an edge in the system. Indeed, by introducing an additional modulation to our chosen $f(t)$ in the form of $ \lambda_b \, e^{-t^2\Omega^2/2\sigma^2}$\changed{, where $\lambda_b=-9.8\times10^{-6}$,} we create a barrier in $\bar\varepsilon(z)$ represented by a Gaussian function with a width of \changed{$\pi/16$.} In this case, we observe the appearance of two levels
in the quasi-frequency spectrum located in the gap if $\lambda_1<0$ [Fig.~\ref{fig:2}(b)]. In the moving frame, the electromagnetic fields corresponding to these levels are localized on either side of the barrier, i.e.~they are topologically protected edge states \cite{Asboth2016short}. In the laboratory frame, if we place a detector at a certain position $z_0$ in the resonator, we can observe these edge states in the time domain \cite{Lustig2018,Giergiel2018,SachaTC2020,Hannaford2022,kopaei2022}. This means, temporal changes of the probability of the detector clicking reveal the appearance of an edge state in time when the edge in $\bar\varepsilon(z_0-\Omega t)$ reaches the detector position. \changed{ Figures~\ref{fig:2}(c)-\ref{fig:2}(d) show} how the electric field changes in time at the detector position for both the edge states and the so-called bulk states. The latter are delocalized in time across the entire period $T=2\pi/\Omega$. 
\changed{In this experiment, a pulse with the central wavenumber $k_0$ and temporal width of $0.119/\Omega$ is injected at a specific position $z$. If the injection occurs precisely when the edge in $\bar\varepsilon(z-\Omega t)$ passes by, \changed{the} pulse will propagate without any distortion, due to the topological properties of the system. In contrast, if the pulse is injected at other times, it will quickly spread over the entire resonator within a characteristic timescale of the order $s/\Omega$. }
Note that the depth of the permittivity modulation needed to realize the described phenomena is very small, \changed{i.e. maximally of the order $10^{-5}$.} 

\begin{figure}[t]
\centering
\includegraphics[width=0.99\columnwidth]{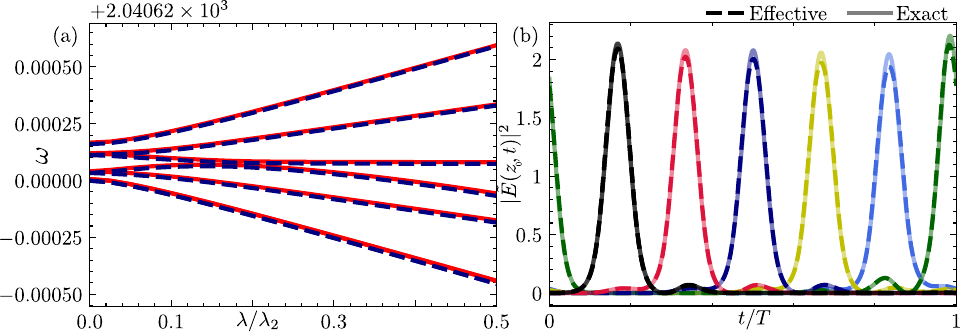}
\caption{(a) Quasi-frequency spectrum as a function of the strength, $\lambda$, of the artificial static electric field generated in (\ref{epsilonSSH}) by the additional time modulation of the permittivity (see text) for $s=6$, $\lambda_1=0$ and \changed{$\lambda_2=10^{-6}$.} (b) In the presence of the artificial static electric field, five solutions of the Maxwell equations exhibit Stark localization which in the laboratory frame we observe in the time domain. Namely, the electromagnetic fields at a fixed position in the laboratory frame \changed{(here $z_0=0$)} are localized at different moments in time. These moments correspond to different local minima of $\bar\varepsilon(z_0-\Omega t)$. There is one more solution (green curve) localized around the discontinuity in $\bar\varepsilon(z_0-\Omega t)$ corresponding to the third quasi-frequency level in (a). The presented states correspond to \changed{$\lambda/\lambda_2=0.5$}.
}
\label{fig:3}
\end{figure}

As a second example, we consider the introduction of an artificial static electric field in the crystalline structure described by (\ref{epsilonSSH}) and predict the observation of the Wannier-Stark localization in the optical system \cite{Wannier_1962,Philipp_2015}. Knowing the Fourier expansion of the linear potential in a resonator, $\lambda (z-\pi)=i\lambda\sum_{m\ne 0} e^{imz}/m$, we can introduce additional Fourier coefficients in $f(t)$, i.e. $f_{-m}=i\lambda/(mh_m)$, which lead to a tilted crystalline structure and thus the presence of an artificial static electric field in our system. Note that in a ring-shaped resonator, the linear potential has a discontinuity at $z=0$ (or equivalently \changed{at} $z=2\pi$). However, away from $z=0$, the influence of the boundary conditions is negligible, and we can observe states that lose the character of extended Bloch waves and localize in different cells of the tilted crystalline structure $\bar{\varepsilon}(z)$. In the case of $s=6$, which is presented in Fig.~\ref{fig:3}, we see five such states with equally spaced quasi-frequencies. There is one more state localized around the discontinuity whose quasi-frequency falls between the second and third equally spaced eigenvalues [Fig.~\ref{fig:3}(a)]. In the laboratory frame, with a detector at a certain position $z_0$ in the resonator, we can observe the electromagnetic field localized at different moments in time \changed{if electromagnetic pulses are injected into the resonator at proper moments} [Fig.~\ref{fig:3}(b)]. 

Different time crystalline structures can be combined by choosing the appropriate time modulation. Importantly, the realization of these structures in the optical regime does not require either deep modulation of the refractive index or modulation frequencies comparable to optical frequencies, as is the case in already known photonic time crystals. Thus, we obtain a tool that allows for processing of electromagnetic signals where time crystalline structures play a key role. \changed{This approach has practical applications and} \changed{introduces a path towards} optical timetronics. 

\changed{ As examples, we considered a resonator in the shape of a one-dimensional ring in this Letter. We quantitatively} examined the necessary conditions to realize the Su–Schrieffer–Heeger and Wannier-Stark ladder models, but our findings extend well beyond these particular cases. \changed{Our idea can be \changed{ readily extended to two-dimensional, and possibly higher-dimensional, }  condensed matter phases implemented in periodically modulated coupled traveling-wave resonators. \cite{SM}.} Notably, the introduction of a non-linear medium, where the permittivity varies with the electric field strength, enables us to introduce interactions \cite{SM}. \changed{If the energy (in frequency units) associated with the nonlinear terms is comparable to or smaller than the bandwidths of the quasi-frequencies shown, for instance, in Figs.~\ref{fig:2}-\ref{fig:3}, then nonlinearity will not destroy the crystalline structures but instead will provide the possibility to obtain qualitatively different solutions within those structures.}
This capability opens up new opportunities to explore and study a wide range of condensed matter phases and phenomena. \changed{Finally, in this work, we focus on time-periodic modulations that preserve time-reversal symmetry.
Our formalism, however, can be extended to model 
\changed{systems with time reversal} symmetry-breaking modulations, resulting in an asymmetric effective permittivity $\bar\varepsilon(z)$. 
This is analogous to quantum systems, where \changed{the presence of magnetic field breaks time-reversal symmetry.}
Such extensions open new opportunities for realizing 
photonic devices and synthetic gauge fields.}

The data that support the findings of this study are available from the corresponding
author upon reasonable request.

{\em Acknowledgment.}~This research was funded by the National Science
Centre, Poland, Project No.~2021/42/A/ST2/00017. The numerical computations in this work were supported in part by PL-Grid Infrastructure, Project No.~PLG/2023/016644. This research was funded in whole or in part by the Austrian Science Fund (FWF) [grant DOI: 10.55776/ESP171]. HT was supported by the National Science Foundation of the United States of America under Grant No.~2131402. Research conducted by AM was carried out at the Jet
Propulsion Laboratory, California Institute of Technology, under a contract with the National Aeronautics and Space Administration (80NM0018D0004). DH acknowledges funding from the UK Engineering and Physical Sciences Research Council EPSRC [grant number EP/W524372/1].
\section{Supplementary Material}
\section{Rotating Frame\label{app:rot}}
We begin with the original form of the Maxwell equations, which govern the behavior of the electric $\vect E$ and magnetic $\vect H$ fields in a medium:

\begin{equation}
\nabla \times \mathbf{E} = -\mu_0 \frac{\partial \mathbf{H}}{\partial t}, \quad
\nabla \times \mathbf{H} = \varepsilon_0 \frac{\partial [\varepsilon(\vect r,t) \mathbf{E} ]}{\partial t}.
\label{eq:1_r}
\end{equation}
Here, $\mu_0$ represents the vacuum permeability, and $\varepsilon_0$ is the vacuum permittivity and $\varepsilon(\vect r,t)$ is the relative permittivity of a medium. 
To facilitate the analysis, we perform a re-scaling of the magnetic field, $\vect H \rightarrow  \vect H\sqrt{\mu_0/\varepsilon_0}$. Moreover, for the ring-shaped resonator we consider in the main text, it is convenient to use $L/2\pi$ and $L/2\pi c$ as the units of length and time, respectively, where $L$ is the circumference of the ring and $c=1/\sqrt{\varepsilon_0\mu_0}$.
This choice of units is particularly convenient for systems with a periodic geometry, as it naturally aligns with the symmetry of the problem. By scaling length by the factor $L/2\pi$, we effectively normalize the spatial dimension to the geometry of the ring, turning the circumference into $2\pi$. Similarly, scaling time by $L/2\pi c$ normalizes temporal dynamics to the time it takes for light to travel around the ring. Under this transformation, we obtain

\begin{equation}
\nabla \times \mathbf{E} = - \frac{\partial \mathbf{H}}{\partial t}, \quad
\nabla \times \mathbf{H} = \frac{\partial [\varepsilon(\vect r,t) \mathbf{E} ]}{\partial t}.
\label{eq:4_r}
\end{equation}

In a ring-shaped resonator with a square cross-section, as depicted in Fig.~1(a) in the main text, we analyze a scenario where the relative permittivity, $\varepsilon(\vect r,t)$ is equaled to a constant value $\varepsilon_r$ except within a localized segment in the resonator. In this segment, the permittivity is subject to periodic modulation over time. Specifically, the permittivity is modeled as 
\be
\varepsilon(\vect r,t) = \varepsilon(z,t) =\varepsilon_r + h(z) f(t),
\ee
where $z$ represents the position along the resonator circumference. The function $h(z)$ is spatially localized, effectively defining the region within the resonator where the permittivity varies in time. On the other hand, $f(t)$ is a periodic function with a period $T$, such that $f(t+T) = f(t)$, describing the time-dependent variation in the permittivity within the localized region, as illustrated in Fig.~1(b) in the main text.

We focus on the TE$_{11}$ mode in the resonator. The appropriate ansatz which satisfies the metallic boundary conditions at $x=0$ or $a$ and $y=0$ or $a$, where $a$ is the length of each side of the cross-sectional area of the resonator [Fig.~1(a)], is given by
\bea
\label{eq:1}
\mathbf{E}&=&  \left[\cos(k_{\perp}x)\sin(k_{\perp}y)\vect{e}_x-\sin(k_{\perp}x)\cos(k_{\perp}y)\vect{e}_y\right]
\cr && \times E(z,t)
\cr 
\mathbf{H} &=& [\sin(k_{\perp}x)\cos(k_{\perp}y)\vect{e}_x+\cos(k_{\perp}x)\sin(k_{\perp}y)\vect{e}_y]
\cr 
&&\times H_{t}(z,t) + H(z,t) \cos(k_{\perp}x) \cos(k_{\perp}y)\vect{e}_z,
\eea
where $E(z,t)$ and $H_t(z,t)$ represent the transverse components of the electric and magnetic fields, respectively, $H(z,t)$ is the longitudinal component of the magnetic field and $k_\perp=\pi/a$. It is worth noting that while a square cross-section is chosen here to simplify calculations, the main results will remain valid for a rectangular cross-sectional shape which is more common in photonic integrated circuits. Furthermore, for dielectric resonators the sinusoidal wave ansatz will change, but again the primary conclusions of the study will not be impacted.

The equation $\nabla\cdot\mathbf{H}=0$ allows us to express $H_t(z,t)$ in terms of the longitudinal component,  
\begin{align}
H_t(z,t) =-\frac{1}{2k_\perp} \frac{\mathrm{d}H(z,t)}{\mathrm{d}z}.
\end{align}
Considering that $\varepsilon(z,t)$ is periodic in time such that $\varepsilon(z,t)=\varepsilon(z,t+T)$, the Floquet theorem \cite{Shirley1965,SachaTC2020} implies that general solutions for the transverse electric and longitudinal magnetic fields can be expressed as superposition of $E(z,t) = \widetilde{E}(z,t) e^{i\omega t}$ and $H(z,t) = \widetilde{H}(z,t) e^{i\omega t}$ where $\widetilde{E}(z,t+T)=\widetilde{E}(z,t)$ and $\widetilde{H}(z,t+T)=\widetilde{H}(z,t)$ and the phase factors are determined by the quasi-frequencies $\omega$. \changed{The latter are} eigenvalues of the generalized eigenvalue problem,
\be
\left[ 
\begin{array}{cc}
-i\partial_t\varepsilon-i\varepsilon\partial_t & \frac{i}{2k_\perp}\partial^2_z-ik_\perp  \\
2ik_\perp & -i\partial_t  
\end{array} 
\right]
\left[ 
\begin{array}{c}
\widetilde E \\
\widetilde H 
\end{array} 
\right]
=
\omega
\left[ 
\begin{array}{cc}
\varepsilon & 0  \\
0 & 1  
\end{array} 
\right]
\left[ 
\begin{array}{c}
\widetilde E  \\
\widetilde H 
\end{array} 
\right],
\label{exactMaxwell_appendix}
\ee
derived from the Maxwell equations (\ref{eq:4_r}), $\nabla\cdot\mathbf{H}=0$ and $\nabla\cdot (\varepsilon(z,t)\mathbf{E})=0$. The permittivity $\varepsilon(z,t)$ is periodic in time and in the ring-shaped resonator, with the circumference of $2\pi$ in the units we use, it also fulfills $\varepsilon(z+2\pi,t)=\varepsilon(z,t)$. Thus, $\widetilde E(z,t)$ and $\widetilde H(z,t)$ fulfill periodic boundary conditions both in time and space and to solve the generalized eigenvalue problem (\ref{exactMaxwell_appendix}), we can expand $\widetilde E(z,t)$ and $\widetilde H(z,t)$ in the basis \changed{
$e^{imz} e^{in\Omega t}$} where $\Omega=2\pi/T$. The resulting matrix-form eigenvalue problem can be solved with standard routines. \changed{Note that the imposed periodic boundary conditions in space lead to a very good description of the resonator presented in Fig.~1(a) if the circumference of the ring $L$ is much greater than the transverse size $a$ which is the case considered in the Letter.}

In the main text we consider resonant driving of the system, i.e. the frequency $\Omega$ of the periodic modulation of the permittivity $\varepsilon(z,t)$ matches the free spectral range of the resonator. In other words we focus on electromagnetic waves with wave numbers close to $k_0$ for which the group velocity (calculated
in the absence of the modulation) $\partial\omega/\partial k|_{k=k_0}=\Omega$
in the units we use. In such a case, we can simplify the description by deriving an effective counterpart of the Maxwell equations (\ref{exactMaxwell_appendix}).
First, let us switch to a reference frame moving with the resonant group velocity, $z'=z-\Omega t$, using the unitary transformation $U=e^{\Omega t \partial_z}$. Multiplying the rows of Eq.~(\ref{exactMaxwell_appendix}) by $U$ and inserting the identity operator $1=U^\dagger U$ leads, e.g. in the case of the second row, to
\be
2ik_{\perp} U\widetilde{E}(z,t) -iU\partial_t[U^{\dagger}U\widetilde{H}(z,t)]=\omega U\widetilde{H}(z,t).
\label{eq:6}
\ee
In the moving frame, we define the fields $\widetilde{E}'(z,t)= U\widetilde{E}(z,t)$ and $\widetilde{H}'(z,t)= U\widetilde{H}(z,t)$ and finally obtain the following eigenvalue equation,
\begin{figure}[t!]
\centering
\includegraphics[width=0.99\columnwidth]{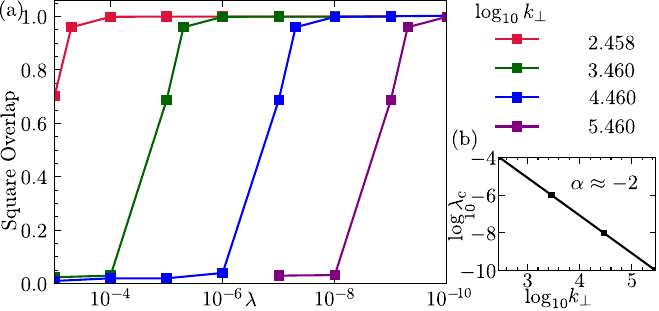}
\caption{
\changed{(a) The square overlap between an eigenvector of the effective equation~(\ref{effMaxwell1}) and the corresponding eigenvector of the full equation~(\ref{roteMaxwell_appendix1}) is shown vs. $\lambda$ for various values of $k_{\perp}$, indicated by different colors. The perturbation in the permittivity is given by $\varepsilon(r,t) = \varepsilon_r + h(z) f(t)$, where $h(z) = \exp(-z^2 / 2\sigma^2)$ with $\sigma = \pi / 41$, and $f(t) = (\lambda / h_{s/2}) \cos(s\Omega t / 2)$ for $s = 12$. (b) displays a log-log plot of the critical modulation strength $\lambda_c$ (defined as the maximal value of $\lambda$ at which the squared overlap is greater than $0.99$) as a function of $k_{\perp}$. The data indicate that $\lambda_c$ scales with $k_\perp$ as $\lambda_c \propto k_{\perp}^{\alpha}$ with the fitted $\alpha\approx -1.998$.
}}
\label{app:fig:2}
\end{figure}
\begin{align}
\label{roteMaxwell_appendix1}
\left[ 
\begin{array}{cc}
-iU\partial_t(\varepsilon U^{\dagger})-iU\varepsilon U^{\dagger}\partial_t & \frac{i}{2k_\perp}\partial^2_z-ik_\perp  \\
2ik_\perp & -i U\partial_t U^{\dagger}-i\partial_t
\end{array} 
\right]
\left[ 
\begin{array}{c}
\widetilde E \\
\widetilde H 
\end{array} 
\right] 
 \\ \nonumber 
=\omega
\left[ 
\begin{array}{cc}
U \varepsilon U^{\dagger}& 0  \\
0 & 1  
\end{array} 
\right]
\left[ 
\begin{array}{c}
\widetilde E  \\
\widetilde H 
\end{array} 
\right],
\end{align}
where we have dropped the primes over the fields to simplify the notation.

If we consider the system with no modulation in time, $\varepsilon(z,t)=\varepsilon_r$, solutions of Eq.~(\ref{roteMaxwell_appendix1}) can be chosen in the form of $\widetilde{E}(z,t)=E_0e^{ikz}$ and $\widetilde{H}(z,t)=H_0e^{ikz}$ with constant $E_0$ and $H_0$ and the corresponding eigenvalue,
\begin{align}
\omega(k) =\sqrt{\frac{1}{\varepsilon_r}(k^2 + 2k^2_{\perp}) }-\Omega k.
\label{disperSM}
\end{align} 
The fields have to fulfill the periodic boundary conditions in the ring-shaped resonator, hence, the dispersion relation (\ref{disperSM}) actually consists of points corresponding to integer values of the wave number $k$.

The dispersion relation (\ref{disperSM}) possesses a minimum at 
\begin{align}
k_{0}= \frac{\sqrt{2\varepsilon_{r}}\Omega k_{\perp}}{\sqrt{1-\varepsilon_{r} \Omega^2}},
\label{minimum}
\end{align}
see Fig.~1 in the main text.
Around this minimum, 
\changed{

\begin{align}
\omega(k) \approx \omega(k_0) +\frac{(k-k_0)^2}{2\sqrt{2\varepsilon_r}k_{\perp}},
\label{approximation}
\end{align}
} 
and the group velocity is very small. Consequently a wave-packet being superposition of waves with the wave numbers $k\approx k_0$ propagates very slowly. Thus, when the modulation of the permittivity is on and it is weak, to describe the system we may average the exact Maxwell equations (\ref{roteMaxwell_appendix1}) over time (rotating wave approximation) and obtain effective time-independent Maxwell equations,
\be
\left[ 
\begin{array}{cc}
i\Omega\partial_z\bar\varepsilon+i\Omega\bar\varepsilon\partial_z & \frac{i}{2k_\perp}\partial^2_z-ik_\perp  \\
2ik_\perp & i\Omega\partial_z  
\end{array} 
\right]
\left[ 
\begin{array}{c}
\widetilde E \\
\widetilde H 
\end{array} 
\right]
=
\omega
\left[ 
\begin{array}{cc}
\bar\varepsilon & 0  \\
0 & 1  
\end{array} 
\right]
\left[ 
\begin{array}{c}
\widetilde E  \\
\widetilde H 
\end{array} 
\right],
\label{effMaxwell1}
\ee
where the averaged permittivity 
\be
\bar\varepsilon(z)=\varepsilon_r+\frac{1}{T}\int_0^T \mathrm{d}t\;h(z+\Omega t)f(t).
\label{effepsilon}
\ee

The comparison between the results obtained from the effective time-independent equations (\ref{effMaxwell1}) and the exact Maxwell equations (\ref{roteMaxwell_appendix1}) for two distinct cases of the permittivity function $\varepsilon(z,t)$ is illustrated in Figs.~2-3 in the main text. These figures clearly demonstrate that the effective approach captures the essential physics of the system while significantly reducing computational complexity and facilitating deeper intuitive insights.

The results presented in Figs.~2–3 in the Letter correspond to a specific resonator geometry, i.e., the ratio of the transverse size to the circumference is $a/L=7\times 10^{-3}$. In an experiment, it may be convenient to use a different geometry to realize a certain temporal crystalline structure. To reproduce the same behavior as, for instance, that described in Figs.2–3, it is sufficient to recalculate the parameters accordingly. In this Letter, we use units in which the ring circumference is always $L = 2\pi$, so a change in geometry amounts to modifying the size of the transverse cross-section $a$. A different $a$ leads to a different value of $k_\perp=\pi/a$, playing the role of effective mass in the dispersion relation \eqref{approximation}. By increasing $k_\perp$ while keeping the same value of the resonant wave number $k_0$ [which implies a corresponding decrease in the modulation frequency $\Omega$, cf. Eq.~\eqref{minimum}], we obtain numerically identical solutions of the effective equations (17) if we reduce the amplitude of the permittivity modulation according to $\lambda \propto 1/k_\perp^2$.

An important question is whether changing the resonator geometry always yields equally good agreement between the solutions of the effective equations and the exact solutions. The results shown in Fig.~\ref{app:fig:2} demonstrate that the {\it critical} modulation amplitude $\lambda_c$ for which, or smaller, we observe nearly perfect agreement between the effective and exact descriptions also scales as $\lambda_c \propto 1/k_\perp^2$. Therefore, having results for a given resonator geometry simultaneously provides the description for a whole class of resonators in which the same phenomenon can be realized.

\section{Possible extension to more than one dimension}

\begin{figure}[t!]
\centering
\includegraphics[width=0.99\columnwidth]{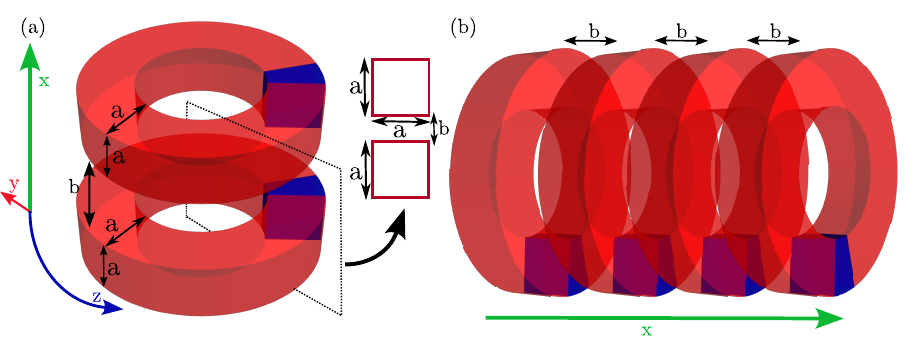}
\caption{
(a) Two resonators in the form of closed rings with square cross-sections. The circumference of the rings, in the units used in the Letter, is $2\pi$. In small segments of the resonators colored by blue, the permittivity is periodically modulated in time with the frequency $\Omega=2\pi/T$. (b) The same setup as in (a) but for a larger number of the resonators, giving rise to two-dimensional space-time structure.}
\label{app:fig:1}
\end{figure}

In this section, we demonstrate how our approach can be generalized to describe two-dimensional crystalline structures. 

Basically, a single ring resonator can support a temporal crystalline structure exhibiting the properties of a one-dimensional condensed matter system. The number of lattice sites in this structure is controlled by the temporal modulation of the permittivity (denoted by the parameter $s$ in our Letter). Now, suppose we have not just a single ring, but many rings arranged in line as shown in Fig.~\ref{app:fig:1} (a)-(b). In each ring, the permittivity can be modulated in the same way allowing the formation of identical temporal crystalline structures. Coupling between the electromagnetic fields propagating in the resonators can occur if the rings are placed sufficiently close to each other. If there are $s$ rings, the system as a whole forms an $s\times s$ two-dimensional crystalline structure, enabling the investigation of two-dimensional condensed matter phenomena. Importantly, the permittivity in each ring can be modulated with a different initial temporal phase, or even in a distinct manner, allowing for the creation of complex, inseparable two-dimensional crystalline structures.

Let us focus on two ring-shaped resonators with square cross-sections as illustrated in Fig.\ref{app:fig:1}a, noting that the generalization to a larger number of rings is straightforward. If the resonators are uncoupled, we obtain a generalized eigenvalue problem in terms of block-diagonal matrices:
\be
\left[ 
\begin{array}{cc}
A_1 & 0  \\
0 & A_2  
\end{array} 
\right]
\left[ 
\begin{array}{c}
\widetilde F_1 \\
\widetilde F_2 
\end{array} 
\right]
=
\omega
\left[ 
\begin{array}{cc}
B_1 & 0  \\
0 & B_1  
\end{array} 
\right]
\left[ 
\begin{array}{c}
\widetilde F_1  \\
\widetilde F_2 
\end{array} 
\right],
\label{2d:exactMaxwell_appendix}
\ee
where $A_1$ and $A_2$ denote the matrix components on the right-hand side of Eq.(\ref{exactMaxwell_appendix}), whereas $B_1$ and $B_2$ correspond to those on the left-hand side. The symbols $\widetilde F_i=[\widetilde E_i, \widetilde H_i]$ stand for the electric and magnetic fields associated with the first and second ring, where $i$ denotes the index corresponding to each resonator and  $ i\in \{1, 2\}$ (see e.g. Fig.\ref{app:fig:1}a). The off-diagonal element of the matrix, denoted as 0, represents a $2 \times 2$ matrix with all zero elements, indicating no coupling between the fields from the first resonator and the second resonator.

To establish coupling between these two waveguides, we adopt the methodology outlined in \cite{haus1984waves,Soltani_16}. In this framework, the electric field of the first waveguide, $\widetilde{E}_1$, couples to the electric field of the second waveguide, $\widetilde{E}_2$, while a similar coupling occurs for the magnetic fields $\widetilde{H}_1$ and $\widetilde{H}_2$. The coupling factor between the waveguides, denoted by $\kappa$, depends on several parameters, including the distance between the waveguides, their material composition, structural design, operating frequency, and the properties of the cladding. 
For simplicity, we treat $\kappa$ as a constant in our calculations. Under this assumption, Eq.~(\ref{2d:exactMaxwell_appendix}) is modified to the following form:
\begin{equation}
\left[ 
\begin{array}{cc}
A_1 & C  \\
-C^* & A_2  
\end{array} 
\right]
\left[ 
\begin{array}{c}
\widetilde{F}_1 \\
\widetilde{F}_2 
\end{array} 
\right]
=
\omega
\left[ 
\begin{array}{cc}
B_1 & 0  \\
0 & B_2  
\end{array} 
\right]
\left[ 
\begin{array}{c}
\widetilde{F}_1  \\
\widetilde{F}_2 
\end{array} 
\right],
\label{2d:exactMaxwell_appendix:coupling}
\end{equation}
where $C$ is a diagonal square matrix with diagonal elements of $i\kappa$ and off-diagonal elements equal to zero.
The matrix elements of $C$ characterize the weak coupling between the resonators, which in turn depends on the distance between them. 

We assume that the coupling between the resonators is weak, i.e. $\kappa$ is at most of the order of the width of the bands presented in Figs.~2-3 of the Letter. It allows us to apply the procedure outlined in the previous section and we obtain a set of coupled two effective equations, each of the form of Eq.(\ref{effMaxwell1}), where the matrices $A_i$ and $B_i$ in Eq.~(\ref{2d:exactMaxwell_appendix:coupling}) are replaced by their time-averaged effective counterparts $A_i^{\mathrm{eff}}$ and $B_i^{\mathrm{eff}}$, respectively.
Due to the fact that elements in the $C$ matrix are time-independent and do not depend on $z$, the effective equation of the entire system will have the simple form of, 
\be
\left[ 
\begin{array}{cc}
A^{\mathrm{eff}}_1 & C  \\
-C^* & A^{\mathrm{eff}}_2  
\end{array} 
\right]
\left[ 
\begin{array}{c}
\widetilde F_1 \\
\widetilde F_2 
\end{array} 
\right]
=
\omega
\left[ 
\begin{array}{cc}
B^{\mathrm{eff}}_1 & 0  \\
0 & B^{\mathrm{eff}}_1  
\end{array} 
\right]
\left[ 
\begin{array}{c}
\widetilde F_1  \\
\widetilde F_2 
\end{array} 
\right].
\label{2d:effectivemax:couple}
\ee

\section{Nonlinear media}
In this section we show how an averaged-permittivity approach can be used to describe a resonator filled with a material which exhibits a Kerr-type cubic nonlinearity. 

In a Kerr-type material, the displacement field is not linear in $\vect{E}$, instead we have
\be
\vect{D}=\varepsilon_0 \left(\varepsilon_1(z,t) + \varepsilon_3 |\vect{E}|^2\right) \vect{E},
\label{eq:nonlinDisp}
\ee
where, $\varepsilon_1$ and $\varepsilon_3$ are the linear and cubic components of the relative permittivity, respectively.

In terms of the fields $E$ and $H$ defined in Eq.~(\ref{eq:1}), Maxwell's equations averaged over the cross-section of the resonator take the form 

\bea
\partial_t [(\varepsilon_1 + \varepsilon_3 |\bar E|^2 ) \bar E(z,t)] &=& \frac{\partial^2_z-2 k^2_\perp}{2k_\perp} \bar H(z,t), \nonumber \\ \cr
\partial_t \bar H(z,t) &=& 2k_\perp \bar E(z,t),
\label{eq:nonlinMaxwell}
\eea
with $\bar E (\bar H)$ denoting space-averaging. In what follows we drop the overbar for brevity. 
Assuming that $\varepsilon_1$ is time-periodic with period $T=2\pi/\Omega$, and that the nonlinearity is weak, Floquet solutions to Eqs.(\ref{eq:nonlinMaxwell}) are stable and can be interpreted as arising from weak perturbations to the linear case, so that the time-dependence of the modulation can be eliminated by time-averaging over solutions in the co-moving frame as in Eqs.(\ref{effMaxwell1}). Applying the unitary transformation, $U=e^{\Omega t \partial_z}$, we obtain

\bea
U\partial_t \left[U^\dagger U\left(\varepsilon_1 + \varepsilon_3 |E|^2 \right)U^\dagger U E\right] &=& \frac{\partial^2_z-2 k^2_\perp}{2k_\perp} U H, \nonumber \\ \cr
U\partial_t U^\dagger U H &=& 2k_\perp U E.
\label{eq:nonlinMaxwellCF}
\eea

At this point we note that the Eqs.(\ref{eq:nonlinMaxwellCF}) are exact up to spatial averaging over the cross section. By assuming that the nonlinearity is weak, we can use a similar argument as in the linear case, wherein the group velocity of solutions in the co-moving frame is small, and they are thus slowly varying. The same applies to $|E|^2$ since it is the envelope of the electric field.
We now perform a time-averaging of 
 Eqs.(\ref{eq:nonlinMaxwellCF}) looking for Floquet solutions of the form $E(H)=\widetilde{E}(\widetilde{H})e^{i\omega t}$ to obtain
\bea
(i\omega + \Omega \partial_z) [\bar\varepsilon_1 + \varepsilon_3 |\widetilde{E}|^2) \widetilde{E}(z,t)] &=& \frac{\partial^2_z-2 k^2_\perp}{2k_\perp} \widetilde{H}(z,t) \nonumber \\ \cr
(i\omega + \Omega \partial_z) \widetilde{H}(z,t) &=& 2k_\perp \widetilde{E}(z,t).
\label{eq:nonlinMaxwellAv}
\eea

Eqs.(\ref{eq:nonlinMaxwellAv}) can now be reduced to a single nonlinear ordinary differential equation which can be numerically integrated by conventional methods. Comparing our system to the nonlinear Schr\"odinger equation for ultra-cold atoms, we can identify $\varepsilon_3$ as an effective scattering length. Similarly to the case of ultra-cold atoms, a wide range of nonlinear condensed matter phenomena can be realized and explored \cite{Dutta2015}.
\bibliography{ref}

\begin{thebibliography}{53}
\bibitem [{foo()}]{footnote1}%
  \BibitemOpen
  \href@noop {} {}\bibinfo {note} {To break the reflection symmetry of the $\bar{\varepsilon}(z)$; we modify the additional modulation to $\lambda_b e^{(t-\gamma)^2/2\sigma^2}$ with $\gamma=0.02$. If such a symmetry was present, the edge states would be superpositions of the localized wavepackets presented in Fig.~\ref{fig:2}(c)}\BibitemShut {NoStop}%
\end{thebibliography}
\end{document}